\documentclass{mpe_report}
\usepackage{psfig}

\begin{document}
\title{X-ray Observations of Galaxies: The Importance of Deep High-Resolution Observations}
\author{G. Fabbiano\inst{1} }  
\institute{Harvard-Smithsonian Center for Astrophysics, 60 Garden St., Cambridge MA, 02138, USA}
\maketitle

\begin{abstract}
X-ray observations of galaxies have grown from a curiosity into a full-fledged field of astronomy. These observations provide unique information on black holes, binary stars, and the hot phase of the ISM, which can be used to constrain the chemical evolution of the Universe, and the joint evolution of galaxies and massive black holes. These exciting results are due in large part  to the high-resolution capability of {\it Chandra}. To follow on {\it Chandra} and push forward this science past the present capabilities, our community must build a high-resolution (sub-arcsecond) large-area (several square meters) X-ray telescope.
\end{abstract}

\section{Why Do We Observe Galaxies in X-rays?}

X-ray observations of galaxies have only been feasible in the past 25 yrs, with the advent of imaging  X-ray telescopes from {\it Einstein} to {\it XMM-Newton}, and in particular with the sub-arcsecond resolution and sensitivity of {\it Chandra} (see e.g. Fabbiano 1989, 2006). Yet, these observations have contributed significantly to our understanding of the nature and evolution of the Universe. We can now observe X-ray binaries (XRBs) and Supernova remnants (SNRs) in a variety of environments and stellar populations,  study of the full gamut of nuclear activity, and investigate the properties of the hot interstellar medium (ISM). 

Combining these results with our knowledge of physics (e.g. the expected spectra of metal-enriched hot plasmas, or models of accretion disk processes) we learn about the astrophysical processes at play. For example, in The Antennae (Baldi et al 2006a, b) detailed measures of the metal abundances of the hot ISM have uncovered a complex metal enrichment and retention picture; soft X-ray images of star-burst galaxies have provided observational evidence of massive outflows  responsible for the  dispersal of these metals in the Universe (e.g. M82; Fabbiano 1988). We have deepened our understanding of accreting sources, XRBs and active galactic nuclei (AGNs), and we have explored a new accretion phenomenon, the ultra-luminous X-ray sources (ULXs). Conversely these observations may increase our knowledge of the laws of physics by providing new information on the states of matter under extreme physical conditions and on the properties of black holes.

This deeper astrophysical understanding, combined with the increasing body of X-ray astronomical observations, has an impact on our understanding of cosmology: the formation and evolution of the Universe and its components. In particular, the detection and study of XRB populations  provide a direct window into the evolution of binary stars, which is in turn linked to the evolution of the stellar component of galaxies; observations of quiet and active galactic nuclei and of their environment in different types of galaxies let us explore the evolution of massive black holes, and the feedback of nuclear activity on the associated galaxy; the study of the metal abundances in hot plasmas and the associated hot outflows provides observational input to the understanding of the chemical evolution of the Universe.

\section{The Need for Deep High Spatial Resolution Observations}

Galaxies are both complex and faint in X-rays. Therefore, observationally, high angular resolution detectors and large collecting areas (or very long exposures with {\it Chandra}) are needed, for both understanding nearby galaxies, and then using this knowledge to unravel the observations of the deep X-ray sky. Here, I offer a few examples to illustrate this statement, based on my own work. This is by no means a complete list, but still makes the point. The first three examples regard issues that have plagued the interpretation of the X-ray data for a couple of decades, and have been solved with {\it Chandra}; the other three examples are all based on what we have learned from very long {\it Chandra} observations that push the limits of this telescope. I will explore the last two examples  in some depth in the following sections. 

\begin{itemize}
\item
Example 1: poor spatial resolution biased the interpretation of spatially complex X-ray spectra towards lower metal content, therefore providing incorrect constraints to the chemical evolution of these systems. This has been proved by comparing the complex {\it Chandra} results with the results of the analysis of the low-resolution {\it ASCA} data (e.g. in The Antennae, see Baldi et al 2006a, b; in elliptical galaxies, see NGC1316, Kim \& Fabbiano 2003).
\item
Example 2: poor spatial resolution resulted in overestimating the extent (and presence) of hot halos in `X-ray faintÕ E and S0 galaxies, resulting in biased mass measurements (all pre-{\it Chandra} results; see e.g. Kim \& Fabbiano 2003, for NGC1316)
\item
Example 3: a correlated effect, the same poor spatial resolution resulted in the debated magnitude of the low-mass X-ray binary (LMXB) contribution  to the X-ray emission of E and S0s, also solved with {\it Chandra} (see refs. in Fabbiano 2006).
\item
Example 4: sub-arcsecond resolution and long exposures are needed to study galaxies at high {\it z}; these galaxies have been detected in stacking experiments (e.g., Brandt et al 2001), but detection of their components cannot be obtained with the present and planned X-ray telescopes. This is essential to separate cleanly the point-like nuclear massive black hole emission, from the more extended integrated emission of the other galactic sources.
\item
Example 5: sub-arcsecond resolution and long exposures are needed to study the entire gamut of nuclear black hole  emission, and explore galaxy-AGN feedback, as demonstrated for example by the {\it Chandra} observation of the quiet nuclear super-massive black hole (SMBH) in NGC821 (see below).
\item
Example 6: sub-arcsecond resolution and long exposures are needed to constrain the evolution of XRB populations in galaxies (as demonstrated by the very deep observations of NGC3379 with {\it Chandra}, see below).
\end{itemize}

\section{Constraining Emission and Fuel of SMBHs: NGC 821}

NGC821 is an elliptical galaxy (D=24 Mpc) with an old stellar population and a nuclear SMBH of 8.7$\times 10^7 M_{\odot}$, corresponding to an Eddington luminosity $L_E \sim 1 \times 10^{46}$erg~s$^{-1}$. However, this nucleus is remarkably inactive (see refs. in Fabbiano et al 2004; Pellegrini et al 2007a). 

Our study of this galaxy demonstrates the need of deep high-resolution observations to explore inactive or weakly active SMBHs. NGC821 was observed with {\it Chandra} first in 2002 for 39 ks (Fabbiano et al 2004); this observation revealed a fuzzy, kpc-size S-shaped central emission, suggestive of a jet or hot filament, a handful of point-like sources (most likely luminous low-mass X-ray binaries - LMXBs, with (0.3-8 keV) $L_X > 1.2 \times 10^{38}$erg~s$^{-1}$), and some circumnuclear diffuse emission, which could be due to hot gas, and therefore be a source of fuel for the SMBH. 

These results were tantalizing enough to grant a deeper look. We now have a total of 230~ks with {\it Chandra} (Pellegrini et al 2007a, b), which have led to the detection of 41 sources (03-8 keV $L_X > 3 \times 10^{37}$erg~s$^{-1}$), with typical LMXB colors and luminosity function. Using globular clusters detected in both the {\it Chandra} and {\it Hubble} images of NGC821 for accurate astrometry, we have identified the nuclear source. This source is a slightly extended, hard ($\Gamma = 1.5$) emission region of $L_X \sim 6 \times 10^{38}$erg~s$^{-1}$, with some elongated (possibly jet-like) emission nearby. 

These deep observations also set a stringent upper limit on the possible amount of  circumnuclear hot ISM; most if not all of the diffuse emission can be explained with unresolved LMXBs. The spectrum of the diffuse emission, after all the detected LMXBs are subtracted, is hard and compatible with that of the LMXBs; moreover, the spatial distribution of this emission follows that of the stellar light and of the number density of detected LMXBs (fig. 1); finally, a comparison of hot and soft bands shows that a soft excess is only marginally possible in the central 10" (and  within the uncertainties it may not be present). The extrapolation of the X-ray luminosity function of detected LMXBs to lower luminosity, and the inclusion of the expected stellar X-ray emission also accounts entirely for any residual diffuse emission.
  
\begin{figure}
\centerline{\psfig{file=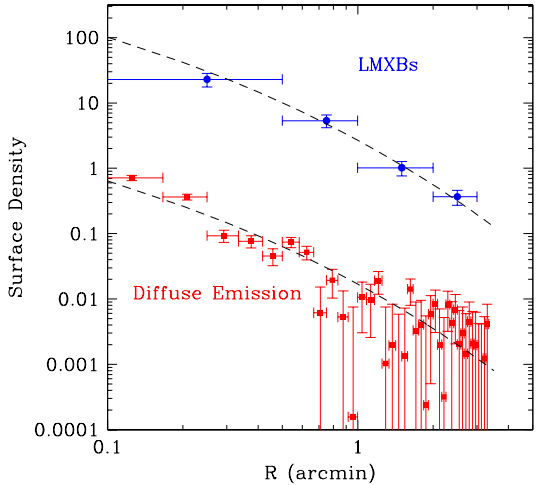,width=8.8cm,clip=} }
\caption{Radial profiles of detected point sources (blue), in units of number per square arcminute; background-subtracted diffuse emission over 0.3Ð6 keV (red), in units of counts per pixel; and galactic R-band emission (dashed lines; arbitrarily normalized).  (from Pellegrini et al 2007a).
\label{image}}
\end{figure}

With these data we can set an upper limit on the density of hot ISM close to the nucleus, based on the fact that any detection of diffuse emission in the short exposure can now be explained away as the result of unresolved LMXBs. This density corresponds (via the Bondi formula) to a mass accretion rate that can produce $L_{bol} \sim 1\times 10^{41}$erg~s$^{-1}$ (for high radiative efficiency), and would be high enough to explain the $L_X$ of the whole nuclear emission. This shows that further high resolution deep observations are needed to either detect or exclude the presence of this hot fuel, and that any coarser look cannot result in meaningful bounds to the nature of the nucleus. Moreover, a different source of fuel for the SMBHs could be colder ISM from stellar outgassing. Again, constraining the hot fuel will result in constraints on the cold fuel, to explain the detected nuclear emission. These results led us to different possibilities for the origin of the nuclear emission, which are beyond the scope of our present discussion, but are highly relevant for the understanding of SMBH Ð feedback in galaxy evolution (see Pellegrini et al 2007a, b).

\section{Understanding LMXB Formation and Evolution: NGC 3379}

LMXBs are the only direct fossil evidence of the formation and evolution of binary stars in the old  stellar populations of early-type galaxies. First discovered in the Milky Way (see Giacconi 1974), these binaries are typically composed of a compact accretor, neutron star or black hole, and a late-type stellar donor. The origin and evolution of Galactic LMXBs has been the subject of much discussion, centered on two main evolutionary paths (see Grindlay 1984; review by Verbunt \& Van den Heuvel 1995):  evolution of  primordial binary systems in the stellar field, or dynamical formation and evolution in globular clusters (CGs). 

With the advent of {\it Chandra}, many LMXB populations have been discovered in early-type galaxies (see review Fabbiano 2006), and the same evolutionary themes (field or GC formation and evolution) have again surfaced, supported and stimulated by a considerably larger and growing body of data. These {\it Chandra} observations  have provided important results on the spatial distributions and X-ray luminosity functions (XLFs) of  LMXB populations (e.g., Kim \& Fabbiano 2004; Gilfanov 2004) and on their association with GCs (e.g. Angelini et al 2001; Kundu et al 2002; White et al 2002). However,  most of the {\it Chandra} observations of LMXB systems consist of fairly shallow individual snapshots for each observed galaxy, with limiting luminosity ($\sim$0.3-8~keV) of a few $10^{37}$~erg~s$^{-1}$. These data give us information on the high luminosity LMXB sources, but do not cover  the typical luminosity range of the  well studied LMXB populations of the Galaxy and M31, which extends down a decade towards dimmer luminosities. 

Apart from rare exceptions, these observations do not have the time sampling that would permit  variability studies and the identification of X-ray transients; both  are  important aspects of the observational characteristics of Galactic LMXBs and are needed for constraining the evolution of these populations (e.g., Piro \& Bildsten 2002, Bildsten \& Deloye 2004). For these reasons we proposed (and were awarded) a very large program of monitoring observations of the two nearby elliptical galaxies NGC 3379 and NGC 4278 with {\it Chandra} ACIS-S3. Here I will discuss some of the NGC 3379 results; we are still working on the NGC 4278 catalog extraction.

NGC 3379, in the nearby poor group Leo (D$\sim 11$~Mpc), was chosen for this study because  is a relatively isolated unperturbed 'typical' elliptical galaxy, with an old stellar populations (age of 9.3~Gyr, Terlevich \& Forbes 2002) and a poor globular cluster system ($S_{GC}=1.3 \pm 0.7$, Harris 1991; where $S_{GC}$ = No.GC$\times 10^{(0.4(M_V+15))}$). These characteristics make NGC 3379 ideal for exploring the evolution of LMXB from primordial field binaries.
Observationally, NGC 3379 is an ideal target for LMXB population studies, because of its proximity, resulting in a resolution of $\sim30$~pc with {\it Chandra}, and the lack of a prominent hot gaseous halo, demonstrated by a previous short {\it Chandra} observation (David et al 2005). These characteristics optimize the detection of fainter LMXBs, and minimize source confusion; because of its angular diameter ($D_{25} = 4.6 $arcmin, RC3), NGC 3379 is entirely contained in the ACIS-S3 CCD chip, and is not affected by the degradation of the {\it Chandra} PSF at large radii. 

NGC 3379 has been observed five times with {\it Chandra} over six years, for a total of 337 ks; the last four observations were all obtained within one year. The coadded data set reaches a detection threshold (0.3-8 keV) $L_X \sim 5 \times 10^{35}$erg~s$^{-1}$, allowing us to study sources in the range of normal Galactic LMXBs. The source catalog derived from these observations (Brassington et al 2007a), reports the detection of 132 sources within or in the immediate outskirts of NGC 3379; 98 of these sources are found within $D_{25}$. Source colors are consistent with those of LMXB populations (e.g., Prestwich et al 2003), and the majority of sources is variable, as expected for compact accreting objects; both flux and spectra vary.

\subsection{Incidence of Transients}
A key observational result that can constrain the nature of these LMXBs is their time variability, because transient behavior is expected from luminous ($L_X \geq 10^{37}$erg~s$^{-1}$) LMXBs if these are relatively detached systems evolved from native X-ray binaries in the stellar field (Verbunt \& Van den Heuvel 1995; Piro \& Bildsten 2002; King 2002). For sources originating as ultra-compact binaries (neutron star - white dwarf) or main-sequence binaries formed in GCs, transient behavior instead is only expected at lower luminosities (Bildsten \& Deloye 2004; Ivanova et al 2007). Our deep monitoring observations of the nearby elliptical galaxy NGC 3379 provide the means to pursue this line of investigation.

While this work is still ongoing, I can report that considering the sources most closely associated with the galaxy (within or very near $D_{25}$) we find 18 transient candidates with on-state luminosity in the $10^{37}$ to a few $10^{38}$erg~s$^{-1}$ range (Brassington et al 2007a, c), three of which are associated with GCs. Since we only have limited time sampling, this result implies a lower limit of 15\% on the number of LMXBs that are transient candidates. I will discuss the GC transients below. Considering only the 15 sources not associated with GCs, which are likely to be dominated by field LMXBs, we find a range of high state duration:  four sources are detected in only one instance, and therefore experiencing short ($< 6$ months) flares; two have on-states lasting in excess of  5 years; seven are on for $> 2$ days; and two are on for $> 4$ months. Of course, the persistent sources could also be transients with on-time $> 5$ yrs. 

\subsection{Low Luminosity Field and GC XLFs}

At luminosities higher than a few $10^{37}$erg~s$^{-1}$ the XLFs of field and GC LMXBs trace each other (Kim E. et al 2006); this result is consistent with a similar origin (GC formation) for the two classes of LMXB, although it does not prove it. Does the similarity extend to the lower luminosities?
 
In NGC 3379, for the first time we can probe deeply the low-luminosity field and GC XLFs of an elliptical galaxy. We find that they differ (Fabbiano et al 2007), in the sense of a highly significant decrease in the number of GCs associated with LMXBs at luminosities $L_X < \sim 4\times10^{37}$ erg s$^{-1}$, compared to field sources, as shown in fig. 2. Using the overlapping {\it Chandra-Hubble WPC3} field of view on this galaxy, we find that the fraction of GCs associated with LMXBs is basically constant at 12-13\% going from an exposure time of 31~ks (Kundu, Maccarone \& Zepf 2007, KMZ) to our 10-fold deeper exposure of 337~ks, while a significant increase in the number of GC-LMXBs would be expected if field and GC LMXBs followed the same luminosity function. A similar effect was suggested by the comparison of GC and Field XLFs in NGC3115, where source detection extends below $10^{37}$ erg s$^{-1}$ (see fig. 6 of KMZ). The study of GC-LMXB populations in M31 and the Milky Way (although distance uncertainties may play a role in the latter) are also consistent with this emerging picture (see figures 6 and 7 of Voss \& Gilfanov 2007a). Our deep look at NGC 3379 suggests that the dearth of low-luminosty GC LMXBs is a general feature of LMXB populations.

\begin{figure}
\centerline{\psfig{file=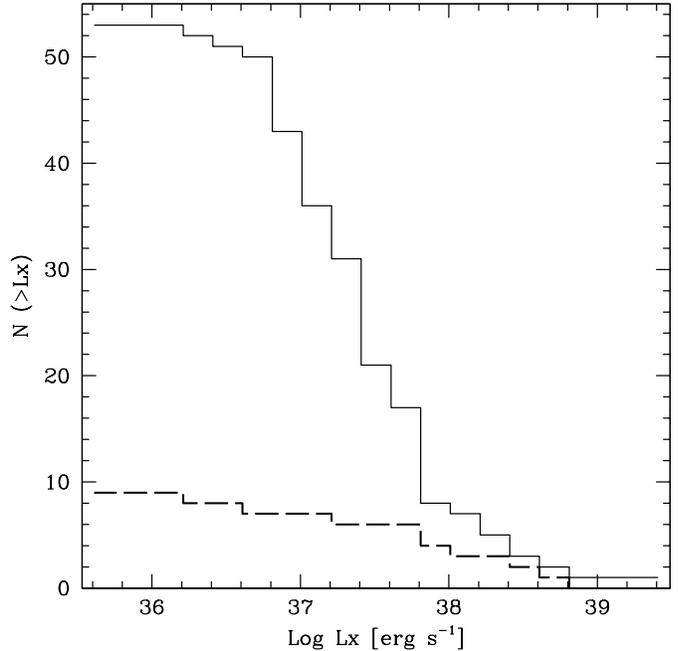,width=8.8cm,clip=} }
\caption{Observed cumulative XLFs of field (continuous line) and GC (dashed) LMXBs; both distributions are from the joint Hubble/Chandra field, and suffer from similar detection biases. A KS test excludes that the two distributions may be derived from the same parent population at 99.82\% confidence (from Fabbiano et al 2007).
\label{image}}
\end{figure}
 
As discussed in Fabbiano et al (2007), the luminosity-dependent differences of field and GC XLFs cannot be explained by high luminosity GCs containing multiple LMXBs, because we find clear evidence of source variability for seven out of the nine GC LMXB sources in NGC 3379, invalidating this hypothesis. Persistent behavior of high luminosity GC sources, compared with transient field sources of similar high luminosity may explain the discrepancy as excess of luminous GC LMXBs. However, the detection of three candidate transient sources (with peak luminosity greater than $10^{37}$ erg s$^{-1}$) in the GC LMXB population of NGC 3379 may not support this explanation. 

The lack of low-luminosity sources in GCs is consistent with the prediction of Bildsten \& Deloye (2004) based on the transition of sources from persistent to transients due to the thermal disk instability. The peak luminosity of two of the three candidate transients we detect is very near $10^{37}$erg~s$^{-1}$, the third has a peak luminosity in the mid-$10^{37}$erg~s$^{-1}$ range.These values could be consistent with the outburst luminosity of a low-luminosity ultra-compact binary, as in the Bilsten \& Deloye model. However, the value of the observed XLF cut-off is not  consistent with their suggestion that GC LMXBs are dominated by ultra-compact binaries, but instead favors LMXBs with H-rich MS donors, which also could be low-luminosity transients (Ivanova et al 2007). The $\sim 4\times10^{37}$ erg s$^{-1}$ luminosity cut-off is also consistent with current theories of magnetic stellar wind braking (Stehle, Kolb \& Ritter 1997), suggesting that this effect may work rather better for the unevolved companions of GC LMXBs than for field LMXBs and cataclysmic variables in the Galaxy, where these companions may be somewhat evolved.

While our results firmly establish a dearth of GC sources in NGC 3379 at low luminosity, the accurate luminosity (and uncertainty) of the GC XLF  cut-off will need the formal analysis of the NGC 3379 LMXB luminosity function (Kim et al in preparation). The forthcoming analysis of the very deep {\it Chandra} observations of NGC 4278, a GC-rich galaxy (see Kim, D.-W. et al 2006), will provide in the near future stronger constraints on the potential 'universality' and value of the GC LMXB XLF cut-off luminosity.

Our results have an important immediate implication for the understanding of LMXB evolution in galaxies: the difference between field and GC LMXB luminosity functions at $L_X < 4\times10^{37}$ erg s$^{-1}$ excludes a single formation mechanism in GCs for {\it all} LMXBs, resolving a long-standing controversy in LMXB formation and evolution (e.g., Grindlay 1984; Grindlay \& Hertz 1985; review by Verbunt \& van den Heuvel 1995). This conclusion is in agreement with the observations of the Sculptor dwarf spheroidal galaxy (Maccarone et al 2005), suggesting that the binary properties of field and GC LMXBs might be different (see KMZ), and with less direct early suggestions, based on LMXB and GC population statistics in elliptical galaxies (Irwin 2005; Juett 2005).

\subsection{Dynamical LMXB Formation in the Inner Bulge?}

If field LMXBs are only the result of the evolution of native field
binaries, we would expect both their number and their spatial
distribution to follow that of their parent population, the old stellar population of the
bulges and early-type galaxies. This was indeed suggested by
correlations between global LMXB luminosity and optical stellar
emission (Gilfanov 2004; Kim \& Fabbiano 2004), and by comparisons of
the radial distributions of the LMXB number density with the optical
surface brightness (e.g. in NGC~1316, Kim \& Fabbiano 2003). The
recent detailed study of the M31 LMXB  population by Voss \& Gilfanov
(2007a, 2007b), has instead questioned this picture, by discovering a
central excess of bulge LMXBs within the central 70\arcsec\ ($\sim$270
pc) radius, over the number expected on the basis of the stellar distribution. These sources appear to correlate with the square of the stellar density, suggesting that they may result from dynamical formation in the inner bulge of M31.

We found no such excess (within radii comparable to those of the bulge of M31 and larger) with our very deep {\it
Chandra} observations of the elliptical galaxy NGC 3379 (Brassington et al 2007b). This suggests a variety of behavior in the dynamical origin for non-GC sources in different environments, which we are exploring in our forthcoming paper. 
Our results are shown in fig. 3, where the source density profile of NGC 3379 is compared with the optical surface brightness distribution. 
  
\begin{figure}
\centerline{\psfig{file=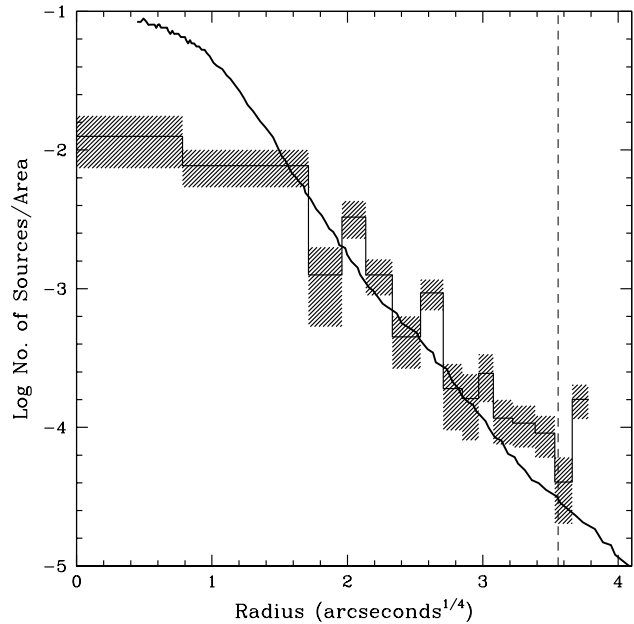,width=8.8cm,clip=} }
\caption{ X-ray source density profile compared to the optical
profile, which indicates the spatial distribution of the stellar
mass. The optical profile (thick line) is the I-band surface brightness best fit of Cappellari et al (2006). This profile was normalized to the X-ray data by way of $\chi^2$ fit.
\label{image}}
\end{figure}

\section{The Future}

How can we progress in the X-ray study of galaxies? Clearly in the next few years we should try to obtain more deep {\it Chandra} observations. But these observations are very expensive, the equivalent of a week or more of observing time. Moreover, {\it Chandra} is a small telescope, so that the amount of photons collected per each source or feature tends to be small. Ten counts may  be a significant detection, but are not enough to study either spectra or variability. It is clear that while we need to maintain or (better) improve the imaging capability of {\it Chandra}, we must increase significantly the collecting area. Only in this way we will be able to survey in depth all the nearby galaxies, study the evolution of galaxies and black holes with redshift, and obtain data that are truly complementary to those that will be obtained by the next generation of space and ground telescopes such as e.g., {\it JWST}, {\it ALMA}, and {\it SKA}. Such a future high-resolution large-area X-ray telescope will require a new approach to X-ray optics. 

A large US-based collaboration successfully proposed a concept for such a mission to NASA as a "Vision Mission" Study in 2004. This very large (0.1-10 keV) X-ray telescope would have 50-100 square meters collecting area and 0.1 arcseconds angular resolution. Based on the success of the Vision Mission Study, we are now proposing to
NASA for a larger, Advanced Mission Concept Study. This study will
refine the mission design and support the development of the detailed
technology plan for the optics and science instruments required for
the next decade. We call this telescope {\it Generation-X}.

\vskip 1.0cm

\begin{acknowledgements}
I particularly thank Nicola Brassington, the postdoc who has done the largest amount of the data analysis in the NGC 3379 project. The recent work presented here is the results of collaborations with several colleagues, including: S. Pellegrini, A. Baldi, D.-W. Kim, R. Soria, V. Kalogera, A. King, S. Zepf, A. Kundu, A. Zezas, G. Trinchieri, J. Gallagher, R. Davies and L. Angelini. I thank R. Brissenden for comments on the {\it Generation-X} mission.
This paper was supported by the {\it Chandra} GO grant G06-7079A (PI: Fabbiano) and sub-contract G06-7079B (PI: Kalogera). Partial support was received from NASA contract NAS8-39073 (CXC).

\end{acknowledgements}
   

\end{document}